\documentclass[12pt]{article}

\newcommand{\bea}{\begin{eqnarray}} 
\newcommand{\eea}{\end{eqnarray}}
\newcommand{\beann}{\begin{eqnarray*}} 
\newcommand{\eeann}{\end{eqnarray*}}
\newcommand{\beq}{\begin{equation}} 
\newcommand{\eeq}{\end{equation}}
  
\newcommand{\6}{\partial } 
\newcommand{\4}{\tilde }
\newcommand{\sfrac}[2]{\mbox{$\frac{{#1}}{{#2}}$}\,}

\begin{document}

\begin{flushright}
AEI-2001-018\\
math-ph/0103006
\end{flushright}

\begin{center}
{\Large
Jet coordinates for local BRST cohomology
}
\end{center}
\begin{center}
{\large Friedemann Brandt}
\end{center}
\begin{center}
Max-Planck-Institut f\"ur Gravitationsphysik
(Albert-Einstein-In\-sti\-tut),\\ 
Am M\"uhlenberg 1, D-14476 Golm, Germany
\end{center}
\begin{abstract}
The construction of appropriate jet space coordinates
for calculating local BRST cohomology groups is discussed.
The relation to tensor calculus is briefly reviewed too.
\end{abstract}


\section*{Introduction}

The BRST formalism has its roots in quantum field theory
where it was first established for gauge theories
of the Yang-Mills type 
\cite{Becchi:1974xu,Becchi:1974md,Becchi:1975nq,
Tyutin:1975qk,Zinn-Justin:1974mc}.
Since then the construction was generalized in several ways.
The best known generalization is the
so-called antifield formalism 
\cite{Batalin:1981jr,Batalin:1983jr}
(for reviews see \cite{Henneaux:1992ig,Gomis:1995he})
which generalizes the
original construction to Lagrangean gauge
theories of any kind (with irreducible or
reducible gauge symmetries, whose 
commutator algebra closes off-shell or
only on-shell). More recently it has been generally proved that
the construction can be generalized so as to include
also global symmetries of the Lagrangian or
`higher order symmetries' \cite{Brandt:1998cz}.
Moreover, the formalism is by no means restricted
to Lagrangean field theories but can be established already
at the level of the equations of motion, whether or not
these equations are Euler-Lagrange equations
\cite{Henneaux:1989cz,Henneaux:1992ig,lecturenotes}.\footnote{Particular 
features of Lagrangean models, such as the 
antibracket \cite{Batalin:1981jr}, have no counterpart in a
general non-Lagrangean model.}
The present work applies to any of these generalizations.

The central object of the BRST formalism
and its generalizations is a
differential $s$ which contains -- 
through the $s$-transformations of the fields
and antifields --
the equations of motion, the nontrivial gauge symmetries (if any),
possibly global or higher order symmetries, as well as related
structure, such as the commutator algebra of the
symmetries contained in it.
The local BRST cohomology concerns cohomology groups of $s$
defined on jet-spaces
associated with the
fields and antifields\footnote{These
are spaces whose coordinates are the coordinates $x^\mu$ of the
base space, the
fields, antifields and derivatives of the fields and 
antifields. In the context of local BRST cohomology it is
useful to count also the differentials $dx^\mu$ among
the jet coordinates. This is done here.};
of particular interest in physics are
$H(s)$ and $H(s|d)$, the cohomology of $s$ and
of $s$ modulo $d$ in the space of
differential forms on these jet spaces, see \cite{Barnich:2000zw} 
for a recent review and for applications.
$H(s|d)$ is related to $H(s)$ through
the so-called descent equations, though the precise relation can
be quite involved \cite{Henneaux:1999rp,Barnich:2000zw}.
Furthermore the physically important cohomology groups
$H^{*,n}(s|d)$ in maximal form-degree $n$ (= base space dimension)
are directly related to
$H(s+d)$ through the 
descent equations 
\cite{Dragon:1996md,Brandt:1997mh,Barnich:2000zw}. 
We denote the combination $s+d$ by $\4s$,
\beq
\4s:=s+d.
\eeq
An important technique within the calculation of $H(s)$ or $H(\4s)$ 
are contracting homotopies which eliminate
pairs of variables from the respective cohomology.
In case of $H(s)$, these pairs are $s$-doublets
$(u^\ell,v^\ell)$,
\beq
su^\ell=v^\ell\quad (\Rightarrow\ sv^\ell=0).
\label{suv}
\eeq
Accordingly they are $\4s$-doublets
$(\4u^{\4\ell},\4v^{\4\ell})$ in case of $H(\4s)$,
\beq
\4s\4u^{\4\ell}=\4v^{\4\ell}\quad (\Rightarrow\ 
\4s\4v^{\4\ell}=0).
\label{tildesuv}
\eeq
However, (\ref{suv}) or (\ref{tildesuv}) alone
do not guarantee the existence of contracting 
homotopies that eliminate the $u$'s and $v$'s, or $\4u$'s and $\4v$'s,
not even locally
(simple counterexamples are the $\4s$-doublets
$(x^\mu,dx^\mu)$ in
Yang-Mills theory, see below).
For (\ref{suv}), a suitable supplementary 
requirement \cite{Brandt:1997mh,Brandt:1999iu} is the existence
of variables $w^I$
which complete the set
$\{u^\ell,v^\ell\}$ to a new jet coordinate
system and are such
that $sw^I$ is a function which can be expressed solely
in terms of the $w$'s,
\beq
sw^I=r^I(w).
\label{sw}
\eeq
Furthermore the $w^I$ are required to be local functions
of the fields and antifields\footnote{The precise definition
of `local function' depends on the context, i.e., on the
model and applications one is studying. Often a local function
is required to live on a finite dimensional jet space,
or to have an expansion in antifields or coupling constants
such that each term in the expansion lives
on a finite dimensional jet space.}.
(\ref{suv}) and (\ref{sw}) imply that $s$ leaves
separately invariant the subspaces of local functions
$f(u,v)$ and $f(w)$ depending only on the $u$'s and
$v$'s, and $w$'s, respectively. 
As an immediate consequence, $H(s)$ factorizes by the
K\"unneth formula (see, e.g., \cite{HiltonStammbach}) 
into the $s$-cohomology
in these subspaces,
\beq
H(s)=H(s,{\cal F}_{u,v})\otimes H(s,{\cal F}_w) ,\quad
{\cal F}_{u,v}=\{f(u,v)\},\quad {\cal F}_w=\{f(w)\}.
\label{kunneth}
\eeq
Hence, $H(s)$ can be obtained by computing separately
$H(s,{\cal F}_{u,v})$ and $H(s,{\cal F}_w)$.
Furthermore (\ref{suv}) implies in many (though not in all)
cases that $H(s,{\cal F}_{u,v})$
is trivial (represented just by constants) so that
the $u$'s and $v$'s disappear from the cohomology and
$H(s)$ is given just by $H(s,{\cal F}_w)$.\footnote{Whether or not
$H(s,{\cal F}_{u,v})$ is trivial depends on the precise properties 
of ${\cal F}_{u,v}$ (in particular on topological features)
and is not to be discussed here
(the focus of the present
work is on the construction of the $w$'s rather
than on contracting homotopies for $u$'s and $v$'s).
An instructive example where $H(s,{\cal F}_{u,v})$
is not trivial can be found in section 5 of \cite{Barnich:1995ap},
see theorem 5.1 there.}

Analogous results hold for $H(\4s)$ if there is
a jet coordinate system
$\{\4u^{\4\ell},\4v^{\4\ell},\4w^{\4I}\}$ such that
\beq
\4s\4w^{\4I}=\4r^{\4I}(\4w).
\label{tildesw}
\eeq

There are three types of $u$'s and $\4u$'s
which one typically meets (see, e.g., the Yang-Mills example
below):
(a) antighost fields and their derivatives,
(b) antifields and a subset of their derivatives,
(c) gauge fields and a subset of their derivatives.
In reducible gauge theories there are normally additional
$u$'s and $\4u$'s given by a subset of 
derivatives of ghosts, ghosts for ghosts etc.
(\ref{sw}) and (\ref{tildesw}) pose no problem in connection with
the antighost fields 
and the corresponding $v$'s
(these $v$'s are Nakanishi-Lautrup fields used for gauge fixing) 
or $\4v$'s
because the $s$-transformations and
$\4s$-transformations  of the other
fields and antifields do not involve these fields%
\footnote{We refer here to the so-called 
``classical basis'' of the antifields. It 
is related to the ``gauge fixed'' basis
simply by a ``canonical transformation'' which is just a local
(anti)field redefinition, see e.g.\ section 2.6 of \cite{Barnich:2000zw}. 
Of course, the cohomology does not
depend on the basis used to compute it.}.
(\ref{sw}) and (\ref{tildesw}) are much more 
challenging -- and much more interesting --
in connection with the $s$-doublets and $\4s$-doublets
associated with the gauge fields and antifields. The construction
of corresponding $w$'s can be quite a 
nontrivial matter. The existence of
such $w$'s is related to a tensor calculus and
a gauge covariant algebra.
This was discussed in \cite{Brandt:1997mh,Brandt:1999iu} 
and will be briefly
reviewed at the end of this letter.

The purpose of this letter is to discuss how and when
one can construct
$w$'s and $\4w$'s fulfilling (\ref{sw}) and (\ref{tildesw}),
respectively.
We shall describe an iterative construction
which can be used as an algorithm
to construct the $w$'s explicitly or to prove their existence
if only one can show that the
iteration terminates or, at least, that it results
in meaningful (local) expressions for the $w$'s
(the algorithm applies to the construction of $\4w$'s as well).
Such a proof is model dependent. To give an example, 
we shall prove that the algorithm terminates
for Yang-Mills theory. We shall also briefly comment on 
how one can master a typical complication in supersymmetric models 
related to the supersymmetry ghosts.

If one can prove in this manner the existence
of the $w$'s, then the algorithm also shows
that the functions $r^I$ in (\ref{sw})
can be already read off from the original $s$-transformations,
i.e., it is not necessary to construct the $w$'s
explicitly in order to determine their $s$-transformations. 
Hence, one may then analyse $H(s)$ even without
having to construct the $w$'s explicitly.
This can have practical use because the explicit
construction of the $w$'s can
be rather involved. 
Furthermore, the functions $r^I$
provide directly the tensor calculus and gauge covariant algebra
associated with the $w$'s and thus
even this algebra can be obtained
without an explicit
construction of the $w$'s.
Analogous comments apply to the functions
$\4r^{\4I}$ in (\ref{tildesw}) and to $H(\4s)$.
The relation to the tensor calculus may be used
in two ways. If an appropriate tensor calculus is
already known for a particular model under study, it may
be useful for the finding or construction of the $w$'s. 
Conversely one may derive a previously unknown
tensor calculus for a model from (\ref{sw}) or (\ref{tildesw}).

\section*{Iterative construction of $w$'s}

Since $s$-doublets are
obvious from the $s$-transformations of the fields and
antifields, our starting point is
the assumption that there is jet coordinate system 
$\{u^\ell,v^\ell,w_{(0)}^I\}$ such that (\ref{suv}) holds
and that $sw_{(0)}^I$ is a power series in the $u$'s and $v$'s,
\beq
sw_{(0)}^I=r^I(w_{(0)})+O(1),
\label{sw0}
\eeq
where $O(1)$ collects all terms that are at least linear in
$u$'s or $v$'s. The aim is to
complete the $w_{(0)}^I$ to local functions $w^I=w_{(0)}^I+O(1)$
which fulfill (\ref{sw}).\footnote{The notation
anticipates that the algorithm yields the same
functions $r^I$ in (\ref{sw}) as in (\ref{sw0}).}
Assume that we constructed already a jet coordinate system 
$\{u^\ell,v^\ell,w_{(m)}^I\}$ with $w_{(m)}^I=w_{(0)}^I+O(1)$ such that
\beq
sw_{(m)}^I=r^I(w_{(m)})+h^I_{m+1}(u,v,w_{(m)})+O(m+2)
\label{swm}
\eeq
where the non-vanishing 
$h^I_{m+1}(u,v,w_{(m)})$ have degree $m+1$ in the $u$'s and
$v$'s and $O(m+2)$ collects all terms of higher degrees.
Note that we refer here to the jet coordinates
$(u,v,w_{(m)})$; expressed in terms
of the coordinates 
$(u,v,w_{(0)})$, $h^I_{m+1}$
has, in general, no definite degree in the $u$'s and $v$'s but
decomposes into terms of various degrees $\geq m+1$, owing to
$w_{(m)}=w_{(0)}+O(1)$.
We now define, again referring to the coordinates
$(u,v,w_{(m)})$,
\beq
\rho_{(m)}:=u^\ell\,\frac{\6}{\6 v^\ell}\ .
\label{rhom}
\eeq
Using (\ref{swm}), one easily computes the
anticommutator of $\rho_{(m)}$ and $s$.
On the $u^\ell$, $v^\ell$ and $w_{(m)}^I$ one gets,
respectively,
\bea
&(\rho_{(m)} s+s\rho_{(m)})u^\ell=u^\ell,\quad
(\rho_{(m)} s+s\rho_{(m)})v^\ell=v^\ell,&
\nonumber\\
&(\rho_{(m)} s+s\rho_{(m)})w_{(m)}^I=
\rho_{(m)} sw_{(m)}^I=Y_{m+1}^I+O(m+2)&
\label{rhouvw}
\eea
where
\beq
Y_{m+1}^I:=\rho_{(m)} h^I_{m+1}(u,v,w_{(m)}).
\label{Y}
\eeq
Note that one has $Y_{m+1}^I=Y_{m+1}^I(u,v,w_{(m)})$
and that $Y_{m+1}^I(u,v,w_{(m)})$ has degree $m+1$ 
in the $u$'s and $v$'s.
Since both $\rho_{(m)}$ and $s$ are antiderivations,
(\ref{rhouvw}) implies, on functions $f(u,v,w_{(m)})$,
\beq
\rho_{(m)} s+s\rho_{(m)}=N_{u,v}+[Y_{m+1}^I+O(m+2)]\,
\frac{\6}{\6 w_{(m)}^I}\ ,
\label{anti}
\eeq
where $N_{u,v}$ is the counting operator for the $u$'s and $v$'s,
\[
N_{u,v}f(u,v,w_{(m)})=
\Big[u^\ell\,\frac{\6}{\6 u^\ell}+v^\ell\,\frac{\6}{\6 v^\ell}\Big]
f(u,v,w_{(m)}).
\]
We now evaluate $(\rho_{(m)} s+s\rho_{(m)})sw_{(m)}^I$ in two ways.
On the one hand, we get, using that $s$ is a differential ($s^2=0$),
\bea
(\rho_{(m)} s+s\rho_{(m)})sw_{(m)}^I&\stackrel{s^2=0}{=}&s\rho_{(m)}sw_{(m)}^I
\nonumber\\
&\stackrel{(\ref{swm})}{=}&
s\rho_{(m)}[r^I(w_{(m)})+h^I_{m+1}(u,v,w_{(m)})+O(m+2)]
\nonumber\\
&\stackrel{(\ref{rhom}),(\ref{Y})}{=}&sY_{m+1}^I+O(m+2).
\label{hand1}
\eea
On the other hand we get, using (\ref{anti}) and 
$N_{u,v}h^I_{m+1}(u,v,w_{(m)})=
(m+1)h^I_{m+1}(u,v,w_{(m)})$,
\beq
(\rho_{(m)} s+s\rho_{(m)})sw_{(m)}^I
\stackrel{(\ref{swm}),(\ref{anti})}{=}
Y_{m+1}^J\,\frac{\6 r^I(w_{(m)})}{\6 w_{(m)}^J}
+(m+1)h^I_{m+1}(u,v,w_{(m)})+O(m+2).
\label{hand2}
\eeq
(\ref{hand1}) and (\ref{hand2}) imply
\[
h^I_{m+1}(u,v,w_{(m)})=\sfrac{1}{m+1}\Big[sY_{m+1}^I-
Y_{m+1}^J\,\frac{\6 r^I(w_{(m)})}{\6 w_{(m)}^J}\Big]+O(m+2).
\]
Using this in (\ref{swm}), the latter yields
\bea
s\Big[w_{(m)}^I-\sfrac 1{m+1}Y_{m+1}^I\Big]&=&r^I(w_{(m)})-
\sfrac{1}{m+1}Y_{m+1}^J\,\frac{\6 r^I(w_{(m)})}{\6 w_{(m)}^J}+O(m+2)
\nonumber\\
&=&r^I(w_{(m)}-\sfrac 1{m+1}Y_{m+1})+O(m+2).
\label{swm+1}
\eea
Defining now
\beq
w_{(m+1)}^I:=w_{(m)}^I-\sfrac 1{m+1}Y^I_{m+1}\ ,
\label{wm+1}
\eeq
equation (\ref{swm+1}) gives
\beq
sw_{(m+1)}^I=r^I(w_{(m+1)})+h_{m+2}^I(u,v,w_{(m+1)})+O(m+3)
\label{recurs}
\eeq
for some $h_{m+2}^I(u,v,w_{(m+1)})$ which has degree
$m+2$ in the $u$'s and $v$'s.
This is an equation as (\ref{swm}), but for $m+1$ in place of $m$.
Furthermore
(\ref{swm}) holds for $m=0$ by assumption (eq.\ (\ref{sw0})).
Hence, the algorithm provides indeed an inductive
existence proof for a set $\{w^I\}$ fulfilling (\ref{sw})
if one can prove
that it terminates or that it results in
closed local expressions for the $w$'s.
Such a proof can often be accomplished by means of appropriate
degrees that can be assigned to the variables, such as the
ghost number and the (mass) dimension of the variables, see
the example below.
Note that the iteration does not modify
the functions $r^I$ and thus one can read off the
resulting functions $r^I$ in (\ref{sw}) already
from the transformations
of the $w_{(0)}$'s in (\ref{sw0}), as promised.

\section*{Yang-Mills theory as an example}

To get an idea how one may prove that the 
algorithm terminates
or results in local expressions
let us assume for a moment that
one can assign a dimension to each of the variables
$u^\ell$, $v^\ell$, $w^I_{(0)}$ such that 
(i) $s$ has dimension 0 when
all (coupling) constants are dimensionless;
(ii) all $u$'s and $v$'s have positive dimensions; 
(iii) there are only finitely many $w_{(0)}$'s with negative
dimensions and all of these variables are Grassmann odd (anticommuting).
Let us also assume that local functions are
power series' in all variables with nonvanishing dimensions
and that the spectrum of dimensions
of the variables is quantized in integer of half-integer units
(which is the standard case).
Then it is easy to prove that the inductive
construction of the $w$'s terminates:
because of (i), the algorithm implies that
each $Y^I_m$ which appears in the construction has the same
dimension as the corresponding $w^I_{(0)}$; 
furthermore, each $Y^I_m$ is a local function
(by assumption, $s$ is a local operator which implies that
$h^I_m$ is local and thus $Y^I_m$ is local too);
because of (ii) and (iii) a local function with definite
dimension cannot
have arbitrarily high degree in the $u$'s and $v$'s
and thus there is a bound $m_I$ for each value of $I$
such that $Y^I_m=0\ \forall m>m_I$ (recall that
$Y^I_m(u,v,w_{(0)})$ contains only terms with 
degrees $\geq m$ in the $u$'s and $v$'s); hence, 
one gets $w^I=w^I_{(0)}-\sum_{m=1}^{m_I}(1/m)Y^I_m$.

An example to which this argument applies
is Yang-Mills theory with standard Lagrangian
\beq
L=-\sfrac 14 g_{ab} F_{\mu\nu}^a F^{\mu\nu b},\quad
F_{\mu\nu}^a=\6_\mu A_\nu^a-\6_\nu A_\mu^a+{f_{bc}}^a A_\mu^b A_\nu^c
\label{YML}
\eeq
where $g_{ab}$ and ${f_{ab}}^c$ are the Cartan-Killing
metric and structure constants of the Lie algebra
of the gauge group, respectively. We denote the
gauge fields and ghost fields by $A_\mu^a$ and $C^a$,
their antifields by $A^{*\mu}_a$ and $C^*_a$, respectively
(the inclusion of antighost fields and Nakanishi-Lautrup fields is
trivial and therefore we neglect them). `Local functions' are
in this example required to depend polynomially on the fields, 
antifields and their derivatives.
The $s$-transformations of these fields and antifields are
\bea
s A_\mu^a &=& D_\mu C^a 
\nonumber\\
s C^a &=&\sfrac 12{f_{bc}}^a C^c C^b
\nonumber\\
s A^{*\mu}_a &=& g_{ab} D_\nu F^{\nu\mu b}
+ C^b {f_{ba}}^c A^{*\mu}_c
\nonumber\\
s C^*_a &=& -D_\mu A^{*\mu}_a+C^b {f_{ba}}^c C^*_c
\label{sYM}
\eea
where $D_\mu$ is the covariant derivative,
\[
D_\mu X^a=\6_\mu X^a+A_\mu^b{f_{bc}}^a X^c\ ,\
D_\mu X_a=\6_\mu X_a-A_\mu^b{f_{ba}}^c X_c\ .
\]
The $s$-transformations of the coordinates and differentials
vanish,
\beq
s x^\mu=0\ ,\ s dx^\mu=0.
\label{sx}
\eeq
The $s$-transformations of derivatives of the fields
and antifields are obtained from (\ref{sYM}) simply through
prolongation, using $[s,\6_\mu]=0$. One verifies
readily that (i) is fulfilled with the following
standard dimension assignments:
\beq
\begin{array}{cc|c|c|c|c}
 & x^\mu,dx^\mu & C^a & A_\mu^a,\6_\mu & A^{*\mu}_a & C^*_a \\
\mathrm{dimension:} & -1 & 0 & 1 & 3 & 4 
\end{array}
\label{dims}
\eeq
Let us now identify $s$-doublets and a corresponding
jet coordinate system $\{u^\ell,
v^\ell,w^I_{(0)}\}$. The antifields
$C^*_a$ and their independent 
derivatives are $u$'s; the corresponding $v$'s,
given by $-\6_\mu A^{*\mu}_a+\dots$ and derivatives thereof,
are taken as elements of the sought jet coordinate system $\{u^\ell,
v^\ell,w^I_{(0)}\}$.
$A^{*\mu}_a$ and
their remaining derivatives of are also $u$'s; their
$v$'s, given by 
$g_{ab}\6_{\nu}(\6^{\nu}A^{\mu b}-\6^{\mu}A^{\nu b})+\dots$ and
derivatives thereof,
are also taken as new jet coordinates.
The remaining $u$'s are the undifferentiated
gauge fields and their symmetrized
derivatives $\6_{(\mu_1}\dots \6_{\mu_k}A_{\nu)}^a$;
the corresponding $v$'s are new jet coordinates substituting for all
derivatives of the ghost fields $C^a$.
A possible set $\{w^I_{(0)}\}$ contains 
thus only the undifferentiated $C^a$,
the $dx^\mu$ and $x^\mu$, and derivatives of the
gauge fields which complete those derivatives
contained already in $\{u^\ell,v^\ell\}$ to a basis
for all derivatives of the gauge fields. 
Property (ii) 
is then fulfilled too. Property (iii) is not fulfilled because
the $x^\mu$ are bosonic variables with negative dimension.
However, this
does not spoil the argument
since the $x^\mu$ are $s$-singlets
which do not occur in the $s$-transformations
of the fields, antifields and their derivatives;
hence, one can simply disregard the $x^\mu$ when applying 
the algorithm and
concludes that the $w$'s exist, without having
to construct them explicitly. 

Of course, since we know the tensor calculus
of Yang-Mills theory, it is not
difficult to find the set of $w$'s explicitly in this case and thus
a formal existence proof is not needed: obviously
$\{w^I\}=\{C^a,dx^\mu,x^\mu,F_{\mu\nu}^a,\dots\}$ fulfills
all requirements
where the nonwritten elements are first and higher
order covariant derivatives of $F_{\mu\nu}^a$
corresponding to the second and higher order derivatives
of $A_\mu^a$ contained in $\{w^I_{(0)}\}$.

However, when one now considers
$\4s$ rather than $s$, then one 
gets already an example where an existence proof 
is simpler than an explicit construction.
Indeed, the existence of $\4w$'s for pure Yang-Mills theory
can be proved along exactly the same lines as the existence
of $w$'s, but the explicit construction
of the $\4w$'s is much more involved than the construction
of the $w$'s.
This is seen as follows. Using that
$\4s$ is dimensionless
with the assignments (\ref{dims}), one readily checks
that all arguments go through for $\4s$ in place of $s$,
with the same $u$'s and $w_{(0)}$'s (i.e.,
one can use $\{\4u^{\4\ell}\}=\{u^\ell\}$ and
$\{\4w^{\4I}_{(0)}\}=\{w^I_{(0)}\}$). One concludes
the existence of a set 
$\{\4w^I\}=\{\4C^a,dx^\mu,x^\mu,\4F_{\mu\nu}^a,\dots\}$
such that (\ref{tildesw}) holds.
The explicit form of these variables is quite involved, though.
For instance,
$\4F_{\mu\nu}^a$ contains $F_{\mu\nu}^a$ but also
antifield dependent terms proportional to
$g^{ab}dx^{}_{[\mu} A^*_{\nu] b}$ and
$g^{ab}dx_\mu dx_\nu C^*_b$, as can be verified by
applying the algorithm explicitly (the origin of these
antifield dependent terms is the presence of
$D^\rho F_{\rho\nu}^a$ in $(s+d)F_{\mu\nu}^a$).

Note that we have counted $x^\mu$ and $dx^\mu$ among
the $\4w$'s in spite of the fact that they form
$\4s$-doublets, 
\beq
\4s x^\mu=dx^\mu.
\label{tildesx}
\eeq
The reason is that the algorithm would not terminate
if one included $x^\mu$ in $\{\4u^{\4\ell}\}$.
The proof sketched above breaks down
because property (ii)
does not hold anymore ($x$ and $dx$ have negative
dimension). What happens, is the following.
$\4s \4w_{(0)}$ contains
$dx^\mu\6_\mu \4w_{(0)}$ and therefore the algorithm
would give, among others,
a contribution $x^\mu\6_\mu \4w_{(0)}$ to
$\4Y_1$ (as $dx^\mu$ would be a $\4v$ when
$x^\mu$ is a $\4u$). This contribution
creates a term $x^\mu x^\nu \6_\mu\6_\nu
\4w_{(0)}$ in $\4Y_2$ and so forth; 
one gets $\4Y_k$ containing
$x^{\mu_1}\dots x^{\mu_k}\6_{\mu_1}\dots\6_{\mu_k}\4w_{(0)}$
with arbitrarily large $k$. 

The Yang-Mills example provides a simple example how
one may keep control of locality. Other models
may be treated similarly, using appropriate
properties that substitute for (i)--(iii).
For instance, when $s$ contains
global or local supersymmetries (with constant
ghosts in the case of
global supersymmetries), the undifferentiated supersymmetry
ghosts are Grassmann even variables with dimension $-1/2$ (as
follows from the supersymmetry algebra, assuming
standard dimension assignments).
Hence, property (iii) is not fulfilled in that case, even when
one disregards $x^\mu$.
Nevertheless, even in supersymmetric models
one may often use arguments similar to those
above by taking the ghost numbers of the variables into
account. For instance, assume that 
all relevant variables with negative
ghost numbers (i.e., the antifields)
have dimensions $\geq 1$ and that the undifferentiated
supersymmetry ghosts and the $x^\mu$ are the only Grassmann
even variables with negative dimensions
(this is the standard case in supersymmetric models, for
the $dx^\mu$ and the translation or diffeomorphism ghosts
are Grassmann odd).
Then one cannot construct local $x$-independent functions
with a fixed ghost number and a fixed dimension containing arbitrarily
high powers of supersymmetry ghosts, i.e., effectively the degree in
variables with negative dimensions is still 
bounded from above.

\section*{Gauge covariant algebras and tensor calculus}

In the following we review very briefly the relation to gauge 
covariant algebras and tensor calculus, referring 
to \cite{Brandt:1997mh,Brandt:1999iu} for details
and further discussion. For illustrative purpose
we assume that there is a jet coordinate system
$\{u^\ell,v^\ell,w^I\}$ with the properties described above,
and that the set of $w$'s contains
variables of ghost number 0 and 1 only,
\beq
\{w^I\}=\{C^M,T^A\},\quad \mathit{gh}(C^M)=1,\quad \mathit{gh}(T^A)=0.
\label{c1}
\eeq
Such a set of $w$'s arises typically in irreducible gauge
theories, such as Yang-Mills theory or gravity (in fact
we have seen above that it arises in pure Yang-Mills theory).
The reason is the following.
An irreducible gauge theory contains only fields with
ghost numbers $\leq 1$ (there are no `ghosts for ghosts').
Assuming that appropriate regularity conditions are fulfilled
(see e.g., \cite{Barnich:1995db,Barnich:2000zw}), all antifields and their
derivatives appear in $s$-doublets.
The same holds of course for the antighost fields.
Hence, all field and antifield variables with negative ghost numbers
will then give rise to $s$-doublets.
Assuming that corresponding $w$'s exist, one gets (\ref{c1}). 
The $C^M$ will be
appropriate ghost variables, typically
corresponding to the undifferentiated
ghost fields and (possibly)
a subset of derivatives of the ghost fields
which do not
appear in $s$-doublets with the gauge fields and derivatives thereof.
The $T^A$, or their antifield independent part, may be interpreted
as tensor fields. 
The typical situation
in reducible gauge theories is similar, the difference being that
then there are also $w$'s with ghost numbers $>1$ corresponding
to ghost fields with higher ghost numbers (`ghosts for ghosts').
In the extended antifield formalism \cite{Brandt:1998cz}, the set $\{C^M\}$
contains in addition constant ghosts corresponding, for instance, to global
symmetries.

Since $s$ has ghost number 1 (i.e.,
it raises the ghost number by one unit),
(\ref{c1}) and (\ref{sw}) imply that
\beq
s T^A = C^M R_M^A(T),\quad
s C^M = \sfrac 12\, (-)^{\epsilon_K+1}C^K C^L {F_{LK}}^M (T),
\label{sT}
\eeq
for some functions $R_M^A(T)$ and ${F_{KL}}^M (T)$ of the $T$'s.
$(\epsilon_K+1)$ denotes the Grassmann parity of $C^K$
and has been introduced to make the following formulae nicer
(it can of course be absorbed into ${F_{KL}}^M (T)$ by redefining
the latter).
Owing just to $s^2T^A=0$ (and thus, in particular,
regardless whether or not the commutator
algebra of the gauge transformations closes off-shell), 
(\ref{sT}) implies an algebra
\beq
[\nabla_M,\nabla_N] = -{F_{MN}}^K (T)\nabla_K
\label{algebra}
\eeq
where $[\ ,\ ]$ is the graded commutator and
$\nabla_M$ are the graded derivations
\beq
\nabla_M=R_M^A(T)\,\frac{\6}{\6 T^A}\ .
\label{nabla}
\eeq
$s^2C^M=0$ gives
the consistency conditions for the algebra (\ref{algebra})
equivalent to
$(-)^{\epsilon_K\epsilon_N}[\nabla_K,[\nabla_M,\nabla_N]]
+\mathrm{cyclic}=0$,
\beq
(-)^{\epsilon_K\epsilon_N}
[\nabla_K{F_{MN}}^L+{F_{MN}}^R {F_{RK}}^L ]
+\mathrm{cyclic}=0.
\label{algebra2}
\eeq
In the Yang-Mills example discussed above, one simply gets:
\beann
\mbox{Yang-Mills:}&&
\{C^M\}\equiv\{C^a\},\quad
\{T^A\}\equiv\{dx^\mu,x^\mu,F_{\mu\nu}^a,\dots\},
\\
&&\{\nabla_M\}\equiv\{\delta_a\},\quad
\{{F_{MN}}^K\}\equiv\{-{f_{ab}}^c\}
\eeann
where $\delta_a$ are the elements of the Lie algebra
of the gauge group in a basis with structure constants ${f_{ab}}^c$.

Similar consequences arise when there is
a jet coordinate system
$\{\4u^{\4\ell},\4v^{\4\ell},\4w^{\4I}\}$ with analogous properties.
The analogue of (\ref{c1}) is
\beq
\{\4w^{\4I}\}=\{\4C^{\4M},\4T^{\4A}\},\quad 
\mathit{tot}(\4C^{\4M})=1,\quad \mathit{tot}(\4T^{\4A})=0
\label{c2}
\eeq
where $\mathit{tot}$ is the degree associated to $\4s$, i.e., the
sum of the ghost number and form-degree,
\[
\mathit{tot}=\mathit{gh}+\mbox{\it form-degree}.
\]
The implications of 
(\ref{c2}) are analogous to (\ref{sT}) through
(\ref{algebra2}). 
However, there is an additional structure
hidden in the resulting equations which comes from the presence of
$d$ in $\4s$. To exhibit it, one decomposes $\4C^{\4M}$
into terms with different antifield numbers. Since
$\4C^{\4M}$ has total degree 1, its antifield
independent piece contains
in general two pieces: 
one piece with form-degree 1 and ghost number 0, and another
one with form-degree 0 and ghost number 1,
\beq
\4C^{\4M}=dx^\mu A_\mu^{\4M}+C^{\4M}+\mbox{terms with
antifields},\quad 
\mathit{gh}(A_\mu^{\4M})=0,\quad
\mathit{gh}(C^{\4M})=1.
\label{tildeC}
\eeq
The antifield independent piece of $\4T^{\4A}$ is denoted
by $T^{\4A}$,
\beq
\4T^{\4A}=T^{\4A}+\mbox{terms with
antifields}.
\label{tildeT}
\eeq
Using this decomposition in 
$\4s\4T^{\4A}=\4C^{\4M}\4R^{\4A}_{\4M}(\4T)$
which is the analogue of the first equation in
(\ref{sT}), one gets by comparing
the coefficients of $dx^\mu$ of the antifield independent
pieces on both sides of the equation,
\beq
\6_\mu T^{\4A}\approx A_\mu^{\4M}\nabla_{\4M} T^{\4A}\ ,\quad
\nabla_{\4M}:=R^{\4A}_{\4M}\,\frac{\6}{\6 T^{\4A}}
\label{partial}
\eeq
where $R^{\4A}_{\4M}$ is the antifield independent part of
$\4R^{\4A}_{\4M}(\4T)$ and $\approx$ denotes equality on-shell
(in general one only has equality on-shell because
the piece in $\4T^{\4A}$ with antifield number 1
gives rise to antifield independent terms in $s\4T^{\4A}$
which vanish on-shell). Normally (\ref{partial}) indicates
that the set $\{\nabla_{\4M}\}$ 
contains covariant derivatives ${\cal D}_m$ corresponding to a subset
$\{A_\mu^m\}$ of $\{A_\mu^{\4M}\}$ which defines an invertible
(generally field dependent) $n\times n$ matrix,
\bea
\{A_\mu^{\4M}\}&=&\{A_\mu^m,A_\mu^{I}\},\quad
\mu,m\in\{1,\dots,n\},
\label{invers}\\
\{\nabla_{\4M}\}&=&\{{\cal D}_m,\nabla_I\},\quad
{\cal D}_m=A^\mu_m (\6_\mu-A_\mu^{I}\nabla_I),\quad
A^\mu_m A_\mu^k=\delta^k_m\ .
\label{D}
\eea
In the Yang-Mills example, one gets:
\beann
\mbox{Yang-Mills:}&&
\{\4C^{\4M}\}\equiv\{dx^\mu,\4C^a\},\quad
\{A_\mu^{\4M}\}\equiv\{\delta_\mu^m,A_\mu^a\},
\\
&&\{\4T^{\4A}\}\equiv\{x^\mu,\4F_{\mu\nu}^a,\dots\},\quad
\{T^{\4A}\}\equiv\{x^\mu,F_{\mu\nu}^a,\dots\},
\\
&&\{\nabla_I\}\equiv\{\delta_a\},\quad
{\cal D}_m=\delta^\mu_m(\6_\mu-A^a_\mu\delta_a).
\eeann

\end{document}